# Mapping Business Process Modeling constructs to Behavior Driven Development Ubiquitous Language

Rogerio Atem de Carvalho, Fernando Luiz de Carvalho e Silva, Rodrigo Soares Manhaes

Emails: ratem@iff.edu.br, fernando.carvalho@iff.edu.br, rmanhaes@iff.edu.br
Nucleo de Pesquisa em Sistemas de Informação (NSI), Instituto Federal Fluminense (IFF),
R. Dr. Siqueira, 273, Campos/RJ, Brazil

**Abstract.** Behavior-Driven Development (BDD) is a specification technique that automatically certifies that all functional requirements are treated properly by source code, through the connection of the textual description of these requirements to automated tests. Given that in some areas, in special Enterprise Information Systems, requirements are identified by Business Process Modeling – which uses graphical notations of the underlying business processes, this paper aims to provide a mapping from the basic constructs that form the most common BPM languages to Behavior Driven Development constructs.

## l. INTRODUCTION

Behavior-Driven Development (BDD) [1] is a specification technique that automatically certifies that all functional requirements are treated properly by source code, through the connection of the textual description of these requirements to automated tests. BDD relays heavily on Test-Driven Development (TDD) [2], which in turn is a technique that consists of writing test cases for any programming task (new or adapted feature, improvements, bug corrections etc), before these implementations are performed. According to Koskela [3], TDD is intended for "solving the right problem right", meaning to achieve the correct solution that exactly matches the business problem.

BDD starts with textual descriptions of the requirements using specific keywords that tag the type of sentence, indicating how the sentence is going to be treated in the subsequent development phases. Although using text may be the more natural way of describing a system's requirements, and besides the fact that BDD provides the structure for doing so, in many cases, in special during the development of Enterprise Information Systems (EIS), Business Process Modeling (BPM) take place, providing models in different notations, such as UML Statechart Diagrams or different Petri Nets flavors.

Therefore, the aim of this paper is to provide a mapping from the basic constructs that form the most common BPM languages to Behavior Driven Development constructs. This paper is a first development of the original ideas presented by [4], and is organized as follows: after this introduction, BDD main features are briefly presented; followed by the focus of this paper, which is to show the mappings of basic BPM languages constructs to BDD constructs; finally some conclusions and future work are presented.

## 2. A Brief Introduction to BDD

This topic aims at introducing BDD, however, since the focus of this paper is on the higher BDD abstractions levels, many details of the BDD process will not be treated here.

BDD starts with the identification of a business requirement using a set of pre-determined tags, forming a simple Ubiquitous Language (UL). The requirement is described according to the following template [5]:

As a **Role**
I request a **Feature**
To gain a **Benefit**

This template is used basically to make the business value of the requirement explicit for developers and users. Following this identification, a series of possible scenarios for the requirement must be written, by describing the scenarios as sets of Given-When-Then constructs:

Given a **Context** (or a system **State**)
When an **Event** happens (or an user **Action**)
Then an **Action** is taken (or a system **Reaction**)

From this point onwards, a tool is used to parse the scenarios and map the natural language sentences into the underlying programming language equivalent calls, while keeping the same abstraction level. As an example, the code below represents the transformation of a scenario into Python code:

Scenario: server is available
  Given there is a resource at "http://localhost:8081/myresource"
  When I request this resource as raw
  Then the response code is 200

After this scenario is parsed, the following calls will be generated:

@step(r'Given there is a resource at "(.*)"')
def given_there_is_a_resource_at_group1(step):
    # code goes here
@step(r'When I request this resource as raw')
def when_i_request_this_resource_as_raw(step):
    # code goes here

```
@step(r'Then the response code is 200')
def then_the_response_code_is_200(step):
    # code goes here
```

With the generated code, the programmer can then write tests that will drive all the design and implementation of the system, by using TDD, which prescribes automated unit tests, pieces of code that excite the code that has to be implemented. By wrapping all implementation code with tests, which in turn are automatically tied to the business requirements using BDD, fulfills the needs for the documentation of requirements in most projects, having the advantage of making the whole system verifiable at any time.

Using BDD allows reducing the risks and effort to implement a given change in an information system; therefore the system can be continuously improved without to fall into the famous Boehm's cost of change curve, which established that the cost of change in a software project increases exponentially through the time [7]. Moreover, the THEN constructs represent the acceptance criteria, giving an objective method to state that a given requirement is "done".

## 3. Mapping from Business Process Representations to BDD

According to [6], the Given-When-Then convention connects the human concept of cause and effect, to the software concept of input/process/output. Also according to [6] this convention "is simply a state transition, and that BDD is really just a way to describe a finite state machine. Clearly "GIVEN" is the current state of the system to be explored. "WHEN" describes an event or stimulus to that system. "THEN" describes the resulting state of the system." Therefore, the set of all scenarios of a given business requirement can be represented by a Finite State Machine, and, as a consequence, as a Petri Net. Although there are lots of business process (or workflow) patterns, as described by [8], they are all formed by basic constructs. Therefore, by mapping these basic constructs, in any notation, to Given-When-Then (GWT) constructs, its is possible to build any of these patterns using BDD Ubiquitous Language. The following subsections will present each construct based on the definitions given by [9][1], which uses representations in YAWL, UML Activities Diagrams, and BPML. In order to avoid repetition, this paper will focus on creating a fourth representation, by using UML Statechart Diagrams. This way, the mappings here presented are valid to the four most used business process representation methods.

---

[1] For the sake of improving readability, this paper will use the pattern definitions given by Fortis and Fortis [9], which are clearer, but equivalent, than the official definitions given by the Workflow Management Coalition.

### 3.1 Sequence Pattern

In a sequence, an activity of the process is enabled only after the completion of the preceding activities and it will enable the following activity of the process [9]. Figure 1 shows this pattern represented in a UML Statechart Diagram, and Code 1 shows its representation in the GWT format.

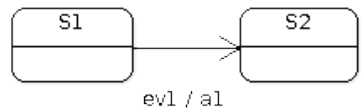

Figure 1: Sequence Pattern

```
GIVEN S1
WHEN ev1
THEN a1
```

Code 1: Sequence Pattern in GWT

### 3.2 Parallel Split Pattern

In parallel split we are referring to a point in the workflow process where a single thread of control splits into multiple threads of control which can be executed in parallel [9].

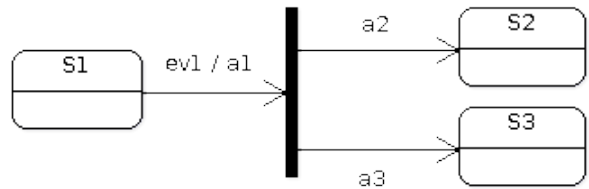

Figure 2: Parallel Split Pattern

```
GIVEN S1
WHEN ev1
THEN a1 AND a2 AND a3
```

Code 2: Parallel Split in GWT

It is important to note that, to the moment, BDD has no "formal" definition for the parallel execution of statements, in these cases, usually only textual information is given in the THEN clauses, leaving to the programmer the task of interpreting them and writing the appropriate code. Another important point is that, although SD can represent "parallel states", it is sometimes confusing to say that "the process is in states E2 and E3". One solution is to create a container state that includes the parallel states, following on the special case of "embedded states" treated in section 3.9.

Also, instead of two THEN clauses, an AND can be used into a single THEN, such as "THEN a5; E2 AND a6; E3."

Code 2 can be understood as "given that the system is in state E1, when the event ev2 occurs then the system should perform actions a5 and a6 and it enters the (composed) state E2 and E3."

### 3.3 Synchronization Pattern

In Synchronization events we observe a point in the workflow process where a two or more thread of control reaches a single thread of control which can then be executed [9].

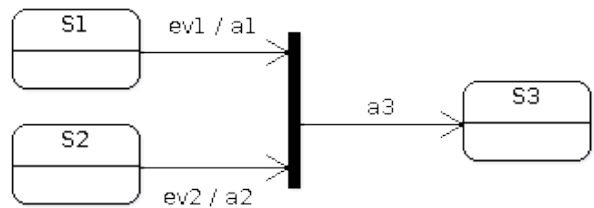

Figure 3: Synchronization Pattern

```
GIVEN S1
WHEN ev1
THEN a1 AND a3

GIVEN S2
WHEN ev2
THEN a2 AND a3
```

Code 3: Synchronization in GWT

### 3.4 Exclusive Choice Pattern

In exclusive choice we are referring to a point in the workflow process where a single thread of control must diverge only to one between multiple threads of control which are available to be executed [9].

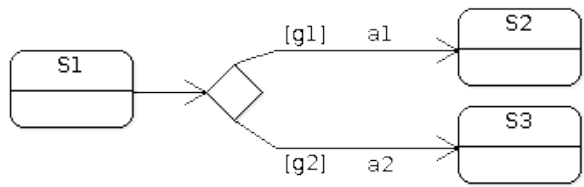

Figure 4: Exclusive Choice Pattern

```
GIVEN S1
WHEN g1
THEN a1

GIVEN S1
WHEN g2
THEN a2
```
Code 4: Exclusive Choice in GWT

### 3.5 Simple Merge Pattern

In parallel split we are referring to a point in the workflow process where a single thread of control splits into multiple threads of control which can be executed in parallel [9].

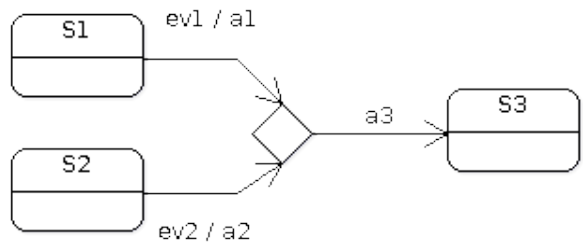

Figure 5: Parallel Split Pattern

```
GIVEN S1
WHEN ev1
THEN a1 AND a3

GIVEN S2
WHEN ev2
THEN a2 AND a3
```
Code 5: Parallel Split in GWT

### 3.6 Multiple Choice Pattern

In multiple choice we are referring to a point in the workflow process where a single thread controls execution or not of multiple threads which can be executed in parallel [9].

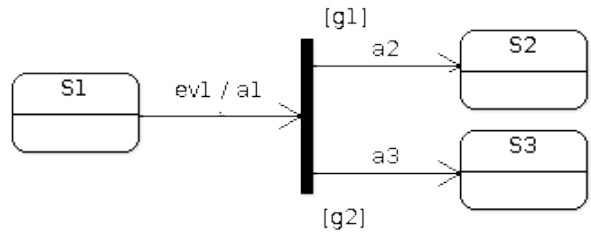

Figure 6: Multiple Choice Pattern

```
GIVEN S1 AND g1 AND NOT g2
WHEN ev1
THEN a1 AND a2

GIVEN S2 AND g2 AND NOT g1
WHEN ev1
THEN a1 AND a3

GIVEN S2 AND g1 AND g1
WHEN ev1
THEN a1 AND a2 AND a3
```
Code 6: Multiple Choice in GWT

### 3.7 Synchronize Merge Pattern

Through this pattern we describe a point in the workflow where multiple paths from the flow converge to a unique thread. If one chooses more than a path, synchronization for the active branches is needed. For a single path choice, alternative branches converge through this point, without synchronization [9]. In statecharts this pattern can be derived from the Synchronization pattern. Statecharts has the advantage of enabling one or multiple branches according to the guard conditions, therefore giving a more expressive representation than Activity Diagrams.

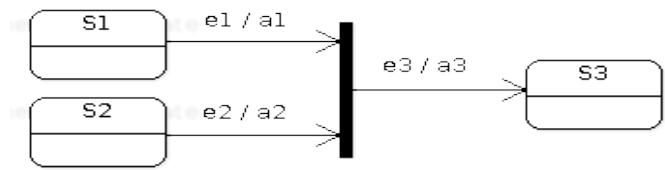

Figure 7: Synchronize Merge Pattern

```
GIVEN S1 AND S2
WHEN e1 AND e2 AND e3
THEN a1; a2; a3 AND S3
```

Code 7: Synchronize Merge in GWT

It is important to note that most of times e3 will not exist, therefore, a3 and the change to the state S3 will happen simply when e1 and e2 occur.

### 3.8 Multiple Merge Pattern

This patterns denominates a point in the workflow in which two or more branches converge, without synchronization. If more than one branch is enabled, possible through concurrent process, the activity following the reunion is launched at any activation of any branch preceding it [9].

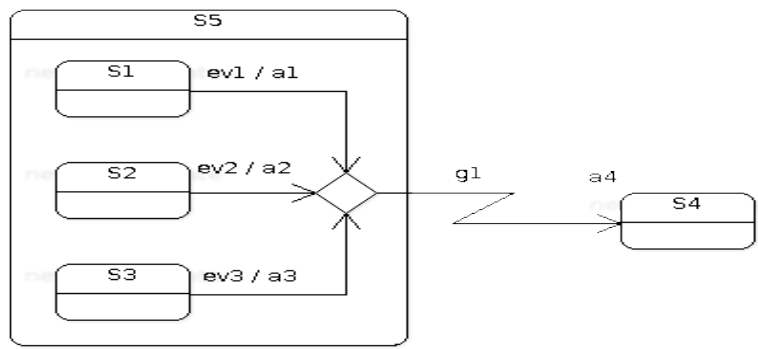

Figure 8: Multiple Merge Pattern

```
GIVEN S1
WHEN ev1 AND g1
THEN a1; a4 AND S4

GIVEN S2
WHEN ev2 AND g1
THEN a2; a4 AND S4

GIVEN S3
WHEN ev3 AND g1
THEN a3; a4 AND S4
```

Code 8: Multiple Merge in GWT

As for the previous pattern, most of times an event such as g1 will not exist.

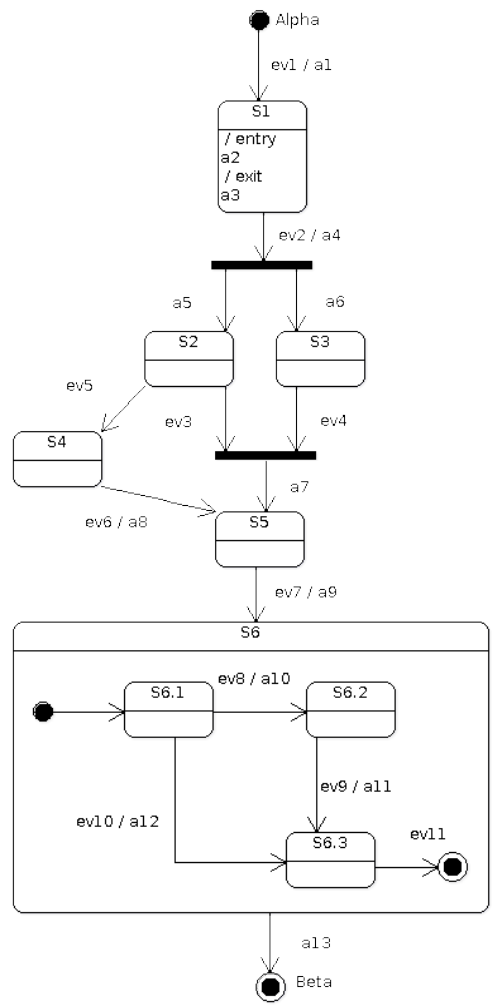

Figure 9: Statechart example with special cases

### 3.9 Special Cases

In this topic we show how to implement additional Statechart Diagrams artifacts that are used to provide more detail to this type of representation. Similarly to the case of parallel actions, to the moment BDD, doesn't provide a "formal" way of describing a given sequence of actions, therefore the listing sequence will be considered in the mappings as the execution sequence.

Figure 9 shows a complex example of a statechart diagram, including some previously treated patterns and the special cases of actions to be performed when the object enters (/entry) or leaves (/exit) the state, and an embedded statechart. Table 1 summarizes the mappings of these special cases to GWT constructs; guard conditions (usually presented between [ ]) are not represented in this example because they are associated to events, already represented previously.

**Table 1: Statechart Diagrams special cases mapped to GWT**

| Case | GWT |
|---|---|
| **/entry** | GIVEN alpha<br>WHEN ev1<br>THEN a1 AND a2 |
| **/exit** | GIVEN S1<br>WHEN ev2<br>THEN a3 AND a4 AND a5 AND S2<br><br>GIVEN S1<br>WHEN ev2<br>THEN a3 AND a4 AND a6 AND S3 |
| **Embedded States** | GIVEN S5<br>WHEN ev7<br>THEN a9 AND S6.1<br><br>GIVEN S6.1<br>WHEN ev8<br>THEN a10 AND S6.2<br><br>GIVEN S6.1<br>WHEN ev10<br>THEN a12 AND S6.3<br><br>GIVEN S6.2<br>WHEN ev9<br>THEN a11 AND S6.3<br><br>GIVEN S6.3<br>WHEN ev11<br>THEN a13 AND Beta |

## 4. Conclusions and Future Work

This paper aimed at describing in more detail the proposal originally presented by [4], which means to associate Business Process Modeling to BDD, by creating a parallel notation of the textual (GWT) constructs by graphical representations of the underlying business process. In that way it is possible to simultaneously benefit from both visual representations of the processes and the security that they are all implemented in the right way through BDD. It is important to note that BDD is a relatively new technique, subject to evolution, therefore, some of the mappings here presented are subject to re-interpretation. Also, specific constructs, such as parallelism, are not explicit in BDD, giving freedom of interpretation for the mappings on one hand, and possible changes of these mappings in the future, on the other hand. Moreover, BDD philosophy is based in simplicity, therefore, some of the constructs here present maybe can be seem as too complex to be notationally represented, or in other words, they must be represented in a textual way.

Through the application of BDD it is possible to reduce risks, costs and effort to implement customizations and changes in general in Enterprise Information Systems such as ERP and CRM. Given an ERP framework wrapped by BDD, it is possible to change business process representations and, by running the associated tests, verify the exact points where the system needs to be changed to make it compliant to the process change. After implementing the changes (and respective tests, of course), it is just a question of running the test set again to check if the new business requirements are correctly implemented.

A series of developments can follow the basic mappings here presented, such as describing them by Formal Methods and implementing the tools originally pointed by [4]:

-Tool for getting a XML representation of the business process and generating the equivalent BDD steps, including the textual representation (GWT) of the process;

-A tool capable of visually running business processes, alternating between its graphical representation and the "living" information system, with step by step execution and definition of paths to follow during the running process. One point of investigation is checking how models@run.time [10] can help implementing this tool and also making the system even more adaptable.

-A translator that, given the GWT constructs, generates the corresponding graphical representation.

These tools should be able to handle the main BPM "languages" such as Petri Nets and UML Statechart and Activity Diagrams.

In an analysis, BDD is Acceptance TDD with a Ubiquitous Language, therefore the next step of this proposal is to provide a complete analysis of the BDD process and map it to a new process that substitutes – in an interchangeable way – the BDD's UL by graphical notations, called Business Process Driven Development (BPDD).